# Spin Injection into a Graphene Thin Film at Room Temperature

Megumi Ohishi, Masashi Shiraishi*, Ryo Nouchi, Takayuki Nozaki, Teruya Shinjo, and Yoshishige Suzuki

*Graduate School of Engineering Science, Osaka University*

*Machikaneyama-cho 1-3, Toyonaka 560-8531, Osaka, Japan*

We demonstrate spin injection into a graphene thin film with high reliability by using non-local magnetoresistance (MR) measurements, in which the electric current path is completely separated from the spin current path. Using these non-local measurements, an obvious MR effect was observed at room temperature; the MR effect was ascribed to magnetization reversal of ferromagnetic electrodes. This result is a direct demonstration of spin injection into a graphene thin film. Furthermore, this is the first report of spin injection into molecules at room temperature.

* E-mail address: shiraishi@mp.es.osaka-u.ac.jp

Graphene,[1] which is an atomically flat layer of carbon atoms densely packed into a benzene-ring structure and in which charge carriers behave as massless Dirac fermions, constitutes a new model system in condensed matter physics. A considerable number of studies on graphene has been implemented, for instance, field-effect transistors with enormously high mobilities,[1] anomalous quantization of the Hall conductance,[2,3] observation of bipolar supercurrent.[4] Although much effort has been invested in studying charge transport in graphene, there have been no reports to date of experimental investigations focused on spin transport in graphene. Since graphene exhibits gate-voltage-controlled carrier conduction and high field-effect mobilities and it consists of only light elements (carbon atoms), which induce a small spin-orbit interaction, it has the potential to become a pivotal material for establishing a new research field of molecular spintronics in which the polarized spin current can be controlled not only by a magnetic field and a bias drain voltage but also by a gate voltage. This, however, is dependent on the ability to inject spins successfully into graphene. There are the other potential materials in non-metallic spintronics, where spin current can be controlled by applying a gate voltage, namely, dilute magnetic semiconductors[5] such as GaMnAs and Si-3d metal alloys.[6] However, spin injection and spin device operation at room temperature (RT) have not yet been successfully demonstrated in either system, which presents a huge obstacle to their application as spintronics materials.

Molecular spintronics has attracted considerable attention recently; in particular, spin-dependent phenomena in π-conjugated organic semiconductors have been studied.[7-16] There have been several important advances in the investigation of spin-dependent phenomena in nanocarbonaceous materials, including spin-dependent transport in carbon nanotubes at low temperature,[7-9] and the observation of a spin-dependent magnetoresistance (MR) effect via $C_{60}$s even at RT.[10-12] It is especially notable that the demonstration of spin injection into carbon nanotubes by introducing a clever experimental technique, namely a *"non-local"* four-terminal measurement.[8] However, in the above-mentioned studies, spin injection into molecules has been definitely demonstrated only at low temperature, while the spin-dependent MR effect at RT has been observed only in $C_{60}$s, where $C_{60}$s behave as tunneling barriers and spins are not injected. In addition, no report of spin injection into other π-electron molecules at RT has been published, as is also the case for nanocarbonaceous materials. Thus, a demonstration of spin injection and spin-dependent transport in nanocarbonaceous materials at RT is critical for the further progress of molecular spintronics and for the development of practical applications.

In this study, we demonstrate for the first time spin injection into a graphene thin film (GTF) with high reliability by employing a *non-local* four-terminal measurement scheme. Of even greater significance, we successfully observed spin-dependent transport at RT. This result opens up the new frontier of non-metallic spintronics and also is the important step toward fabricating molecular spin devices, such as spin field-effect transistors.

Graphene spin devices were fabricated by the following process. The source material of a GTF is highly oriented pyrolytic graphite (HOPG, NT-MDT Co.). We peeled flakes of graphite off by using scotch tape,[1] and then pushed these flakes onto the surface of a $SiO_2$/Si substrate having predefined markers, in which the $SiO_2$ layer was 200-nm thick. When the scotch tape was removed, a GTF was absorbed onto the $SiO_2$ surface by van der Waals force. It should be noted that the absorption force between the GTF and the SiO$_2$ was sufficiently strong that the GTF did not detach from the substrate in the course of the subsequent processes. Non-magnetic (NM) electrodes (Au/Cr=50/10 nm) and ferromagnetic (FM) electrodes (Co=60 nm) were then fabricated onto the GTF using electron beam lithography. The geometry of the two FM electrodes was different in order to generate different coercive forces between the two FM electrodes (see Fig. 1(a)). Figures 1(a) and (b) show a schematic image and a scanning electron microscope image of our graphene spin device; the widths of the two FM electrodes were 700 and 1000 nm, the gap width between the FM electrodes was 250 nm and the contact area between the Co and the GTF was approximately 3 μm$^2$. The thickness of the GTF was estimated to be 10 nm from the atomic force microscope observations.

All MR measurements were preformed at RT by using a four-terminal probing system (JANIS, ST-500) with an electromagnet. The magnetic field was swept from -40 mT to +40 mT in 0.2-mT steps at a sweeping rate of 0.4 mT/s. We employed a standard ac lock-in technique (maximum current = 200 μA, lock-in frequency = 216 Hz, time constant = 300 ms) to measure the MR effect. Detailed experimental schemes of the two-terminal "local" and four-terminal "non-local" measurements are shown in Figs. 2(a) and (b), respectively. The two-terminal local resistance between the NM electrodes was measured to be 77 Ω, that between the FM electrodes was 190 Ω, and the bulk resistance of the GTF was about 5 Ω.

The upper panel of Fig. 3 shows an MR effect observed in the local measurement; the resistance, $R_{local}$, exhibited hysteresis at around -25, -12, +12 and +25 mT. In these measurements, the same measurements were repeated eight times and the data were averaged in order to obtain a higher S/N ratio. Although this result can be explained in terms of an MR effect induced by spin injection into the GTF, this may not actually be the case. One curiosity is the fact that there are in total four hysteresis, although there should be two if the signals would be induced by spin injection. The problem in interpretation of this observation lies in the fact that the other phenomenon, which is not related with spin accumulation and injection into the GTF, influences the MR. As is widely known, injecting a magnetic domain wall into a ferromagnetic nanowire induces a reduction in the resistance due to the anisotropic magnetoresistance (AMR) effect. We used two FM electrodes with different coercive forces in our experiment, and thus the observed four hysteresis can be accounted for in terms of domain wall injection during the magnetization reversal of two FM electrodes having different coercive forces.

In order to exclude these unexpected signals and to isolate spin injection signals from spurious signals, four-terminal non-local MR measurements were introduced. In these measurements, the charge current path and the spin current path are completely separated since the charge current flows anisotropically, while the spin current flows isotropically.[17] The lower panel of Fig. 3 shows the non-local signals, $V_{\text{non-local}}$, measured in the GTF (injected current $I = 100$ μA; the result shown is the averaged data from five measurements); this result is a direct demonstration of spin injection into the GTF at RT. Comparison with the results of the local measurements clearly shows that the change in the non-local resistance occurs in the antiparallel alignment of the magnetization of the FM electrodes. The variation in the signal, $\Delta V_{\text{non-local}}$, decreased linearly from 550 nV to 30 nV, when the injected electric current was reduced from 200 μA to 10 μA, which suggests that the graphene is diffusive. If it is assumed that the spin asymmetry of the resistivity at the Co/graphene interface is the same as the spin polarization of Co, the spin coherence length is comparatively short and is estimated to be about a half of the gap between the FM electrodes (~100 nm). In our study, the non-local MR effect was observed, although the local MR was not. This is attributable to the contact resistance at one Co/graphene interface in our graphene spin device being ~90 Ω, which is approximately 20 times greater than that of the resistance of the GTF. If the spin coherence length is assumed to be 100 nm, as mentioned previously, the expected MR ratio can be easily estimated,[18] and in this case it is calculated to be less than 0.01% (the variation in $R_{local}$ is less than 0.018 Ω, which is smaller than the observed resistance change in the AMR effect). Hence, it is difficult to distinguish the AMR signals and the spin injection signals in case of the local measurements.

In summary, we have succeeded in injecting spins into a GTF at RT, which has not been achieved in previous molecular spintronics and semiconductor spintronics studies. The injection of spins was verified by performing non-local measurements in graphene spin devices. This result has great significance for future developments in molecular spintronics and also opens a door for practical fabrication of three-terminal spin devices, in which a spin current can be modulated by applying a gate voltage.

# Figure captions

**Fig. 1**

(a) A schematic of a graphene spin device. *B* is the external magnetic field.

(b) A scanning electron microscope image of a graphene spin device.

**Fig. 2**

(a) Geometry used for local measurements, in which FM electrodes are used both inject current and measure the voltage.

(b) Geometry of a non-local measurement, in which the voltage circuit is completely separated from the current circuit.

**Fig. 3**

Result of a local MR measurement (upper panel) and a non-local MR measurement at RT (lower panel). The position of the hysteresis in the non-local measurement agrees well with that in the local measurement, which indicates that the hysteresis is induced by spin injection and magnetization reversal of the FM electrodes.

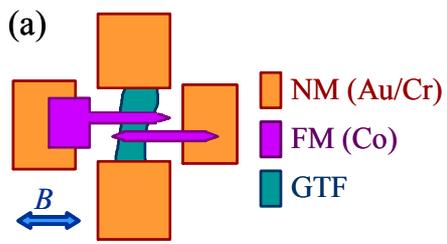
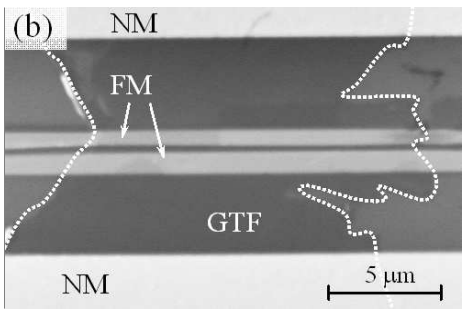

Fig. 1 M. Ohishi et al.,

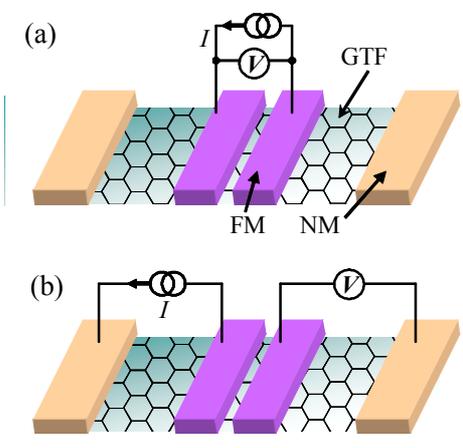

**Fig. 2** M. Ohishi et al.,

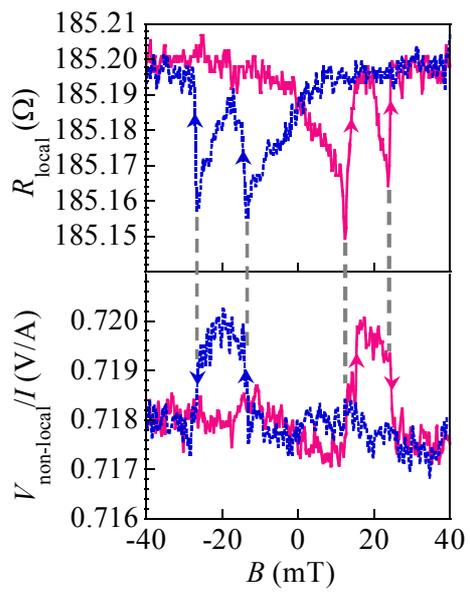

**Fig. 3** M. Ohishi et al.,